\documentstyle[prl,aps,eqsecnum,epsfig,multicol]{revtex}

\begin{document}
\draft

\title{Net Charge Fluctuations in Au + Au Interactions \\
at $\sqrt{s_{NN}}$ = 130 GeV}

\author{
K.~Adcox,$^{40}$
S.{\,}S.~Adler,$^{3}$
N.{\,}N.~Ajitanand,$^{27}$
Y.~Akiba,$^{14}$
J.~Alexander,$^{27}$
L.~Aphecetche,$^{34}$
Y.~Arai,$^{14}$
S.{\,}
% ...
\\(PHENIX Collaboration)\\
}
\address{
% ...
$^{3}$Brookhaven National Laboratory, Upton, NY 11973-5000, USA\\
% ...
$^{14}$KEK, High Energy Accelerator Research Organization, Tsukuba-shi, Ibaraki-ken 305-0801, Japan\\
% ...
$^{27}$Chemistry Department, State University of New York - Stony Brook, Stony Brook, NY 11794, USA\\
% ...
$^{34}$SUBATECH (Ecole des Mines de Nantes, IN2P3/CNRS, Universite de Nantes) BP 20722 - 44307, Nantes-cedex 3, France\\
% ...
$^{40}$Vanderbilt University, Nashville, TN 37235, USA\\
% ...
}

\date{\today}        
\maketitle
\begin{abstract}
Data from Au + Au interactions at $\sqrt{s_{NN}}$ = 130 GeV, obtained with the PHENIX detector at RHIC, are used to investigate local net charge fluctuations among particles produced near mid-rapidity. According to recent suggestions, such fluctuations may carry information from the Quark Gluon Plasma. This analysis shows that the fluctuations are dominated by a stochastic distribution of particles, but are also sensitive to other effects, like global charge conservation and resonance decays.
\end{abstract}


\begin{multicols}{2}   
\narrowtext            

The PHENIX detector~\cite{PHENIX} at the Relativistic Heavy-Ion Collider (RHIC) is a
versatile detector designed to study the properties of nuclear matter at extreme
temperatures and energy densities, obtained in central heavy-ion collisions at
ultra-relativistic energies. A central goal of these studies is to collect evidence
for the existence of the Quark-Gluon Plasma (QGP)  characterized by deconfined quarks
and gluons.

There are several proposed ways to experimentally verify the existence of a
QGP~\cite{QGP}. A general problem is that many of these signals also can be produced
in a hadronic scenario, albeit special conditions of highly compressed matter have to
prevail. Furthermore, it is not straightforward to determine how the various plasma
signals are distorted when the deconfined matter transforms back to hadronic matter.  
Recent theoretical investigations~\cite{Jeon,Asakawa,Heiselberg} predict a drastic
decrease of the event-by-event fluctuations of the net charge in local phase-space
regions as a signature of the plasma state. These fluctuations are not related to the
transition itself, but rather with the charge distribution in the primordial plasma
state. The basic idea is that each of the charge carriers in the plasma carries less
charge than the charge carriers in ordinary hadronic matter. The charge will thus be
more evenly distributed in a plasma. The main concern of the theoretical discussions
is how and why the original distribution survives the transition back to ordinary
matter~\cite{Shuryak,Fialkowski}. Predictions, for a rapidity coverage $\Delta y \geq
1$, range up to an 80\% reduction in the magnitude of the fluctuations, as measured by
the variance of the net charge.

Decays of hadronic resonances influence the net charge fluctuations, whether or not
deconfinement is reached.  In the absence of a QGP, deviations from statistical
behaviour can be used to determine the abundance of e.g.  $\rho$ and $\omega$
mesons~\cite{Jeon2}. In a hadron gas resonances are expected to decrease the
fluctuations by about 25\%.  Globally, fluctuations will be further reduced, since
charge is a conserved quantity. Deviations from a statistical behaviour in the net
charge are also of relevance when the phase transition is treated as a semiclassical
decay of a Polyakov loop condensate~\cite{Dumitru}. Although multiplicity fluctuations
have been studied extensively in both hadronic and nuclear processes~\cite{Wolf}, net
charge fluctuations have not been addressed experimentally.

In this Letter we report results from an analysis of net charge fluctuations for
particles produced in Au+Au interactions at $\sqrt{s_{NN}}$ = 130 GeV. The
fluctuations are studied in the variables $R = n_+/n_-$, the ratio between positive
and negative particles, and $Q = n_+ - n_-$, the net charge~\cite{Jeon}. The
advantages and disadvantages of these variables will also be discussed.

Information from one of the PHENIX central tracking arms (west) is used in this
analysis, where events are required to have a well-defined vertex close to the center
of the apparatus ($|Z|<$~17~cm), as defined by the two beam-beam counters (BBC). These
are Cherenkov-counters surrounding the beam, placed on both sides 1.44 m from the
interaction region, covering the pseudorapidity region 3.0~$< |\eta| <$~3.9. Together
with the information from the two zero-degree calorimeters (ZDC), placed further away
(18~m), the BBC information is used for off-line centrality selection~\cite{Mult}. A
total of about 5$\times$10$^5$ minimum bias events has been analyzed. The PHENIX west
arm spectrometer has an acceptance of 0.7 units of pseudorapidity (-0.35 $< \eta <$
0.35) and $\pi/2$ radians in azimuth $\varphi$. Charged-particle trajectories are
recorded in a multiwire focusing drift chamber (DC)~\cite{general}. The combination of
reconstructed DC tracks~\cite{jeff} with matching hits in the innermost pad-chamber
plane (PC1) defines the sign of the charge of the particle and also provides a high
resolution measurement of the transverse momentum $p_T$ of tracks originating from the
collision vertex. Tracks with a reconstructed $p_T$ less than 0.2 GeV/c have been
excluded from the analysis due to a low reconstruction efficiency and large
contributions from background sources, as revealed by simulations.

The tracking efficiency and the charge assignment have been studied using
GEANT~\cite{GEANT} simulations.  Of particular importance in this context is a
realistic description of the drift chamber response. The drift distances have been
calculated based on a geometric model of the individual drift cells. Additional
parameters describing the response of the chamber, i.e. single-wire efficiency, pulse
width, single-hit resolution, and space drift-time relation, have been extracted from
measured data, parameterized, and applied empirically in the simulation.

RQMD~\cite{RQMD} simulations are used to study the detection efficiency, and the
fraction of reconstructed particles that preserve their charge, as well as to evaluate
the results of the analysis. The charge fluctuations in RQMD are consistent with
calculations based on other hadronic models like UrQMD and HIJING~\cite{Jeon}. The
overall efficiency for detecting a charged particle within the acceptance is found to
be around 80\% for both positive and negative particles.  Depending on $p_T$, between
70\% and 85\% of the reconstructed tracks are in one-to-one correspondence with a
primarily produced particle. The remaining tracks come from secondary interactions in
the detector material and from decays, where the original charge information is lost.

In each event the numbers of positively charged particles $n_+$, negatively charged
particles $n_-$, and their sum $n_{ch}$ are recorded. In a stochastic scenario, with a
fixed number of charged particles within the acceptance, where each particle is
assigned a random charge ($+1$ or $-1$ with the same probability), the variance of the
net charge, Q, is \begin{equation} V(Q) \equiv
{\langle}Q^2{\rangle}-{\langle}Q{\rangle}^2=n_{ch}. \label{e1} \end{equation}
\noindent The normalized variance in $Q$ is \begin{equation} v(Q) \equiv
\frac{V(Q)}{n_{ch}} = 1. \label{e2} \end{equation} For the charge ratio, in the
stochastic scenario, $V(R) \equiv {\langle}R^2{\rangle}-{\langle}R{\rangle}^2$ will
approach the value $4/n_{ch}$ as $n_{ch}$ increases and $v(R) \equiv n_{ch}\cdot V(R)$
asymptotically approaches 4. A small asymmetry between positive and negative particles
affects $v(R)$ drastically, whereas the effect on $v(Q)$ is negligible. If we write
the probability $p_+$, that a given particle has positive charge, in the form $p_+ =\
^1\!/_2(1 + \varepsilon)$, and subsequently $p_- =\ ^1\!/_2(1 - \varepsilon)$, we find
\begin{equation} v(Q) = 1 - \varepsilon^2, \label{e3} \end{equation} while the
asymptotic value of $v(R)$ is $4 + 16\varepsilon + {O}(\varepsilon^{2})$. Detector or
reconstruction inefficiencies do not influence those results in the stochastic
scenario. To calculate $v(Q)$ as a function of multiplicity, Eq.~\ref{e3} can be used
and $v(R)$ can be calculated from \begin{equation} {\langle}R{\rangle} =
\frac{1}{A}\sum_{i=1}^{n_{ch}-1} \frac{n_{ch}-i}{i} \left( \begin{array}{c} n_{ch}\\i
\end{array} \right) \ p_+^{n_{ch}-i} \ p_-^i, \label{e4} \end{equation} and
\begin{equation} {\langle}R^2{\rangle} = \frac{1}{A}\sum_{i=1}^{n_{ch}-1} \left(
\frac{n_{ch}-i}{i} \right )^2 \left( \begin{array}{c} n_{ch}\\i \end{array} \right) \
p_+^{n_{ch}-i} \ p_-^i, \label{e5} \end{equation} where $A = 1 - \ p_+^{n_{ch}} -
p_-^{n_{ch}}$ is the new normalization due to discarding events with $n_+$ or $n_-$
equal to zero. The variance of $R$, even for a purely stochastic charge distribution,
depends on multiplicity and on the fractions of positive and negative particles.

The data show a small excess of positive particles, growing proportionally with
$n_{ch}$, in qualitative agreement with calculations using RQMD and GEANT. A part of
this excess comes from the intrinsic isospin asymmetry and a part from secondary
interactions in the detector and surrounding materials.

In Fig. 1a, $v(R)$ and $v(Q)$ are displayed as functions of $n_{ch}$. $v(Q)$ is
multiplied by a factor of 4 to compensate for the asymptotical difference between v(R)
and v(Q).  Both $v(Q)$ and $v(R)$ are well described by the results obtained from the
stochastic scenario, including the positive excess, as given by the curves.

Since $v(Q)$ is independent of $n_{ch}$ one expects $v(Q)$ to be close to unity also
in representations where other centrality measures are used. On the other hand, since
$v(R)$ depends on multiplicity, it will have a complicated behaviour as a function of
centrality, making it difficult to draw any further conclusions. We will thus focus on
$v(Q)$ for the rest of this analysis.
 
In Fig. 1b, $v(Q)$ is displayed as a function of centrality based on the ZDC/BBC
information. The full event sample, corresponding to 92\% of the inelastic cross
section~\cite{Mult}, is divided into 20 centrality classes, where each class
corresponds to 5\% of the events. Class 20 represents the most central events. With
the increased resolution on the y-axis in Fig. 1b, it is evident that $v(Q)$ is
consistently below unity, and deviates from stochastic behaviour. The value is,
however, far above the most optimistic QGP predictions $v(Q) \sim 0.2$~\cite{Jeon},
although one should keep in mind that our coverage in rapidity is on the limit for
these predictions and that we have only partial coverage in azimuth.

There may be other explanations for deviations from stochastic behaviour than the one
offered by the quark-gluon plasma. These include global charge conservation and
neutral resonances decaying into correlated pairs of one positive and one negative
particle. Both of these effects will decrease the fluctuations, and the decrease will
grow in proportion to the experimental acceptance. In a stochastic scenario, taking
global charge conservation into account, the normalized variance $v(Q)$ becomes
$(1-p)$, where $p$ is the fraction of observed charged particles among all charged
particles in the event.  Eventually, if all charged particles are detected, $v(Q)$
will become $0$.

Experimentally we can change the fraction $p$ of particles within the acceptance by
using different regions of the PHENIX west arm. In Fig. 2, $v(Q)$, for the 10 \% most
central events, is displayed as a function of $\Delta\varphi_d$, i.e. the chosen
azimuthal interval of the spectrometer. For comparison, the results from RQMD
processed through GEANT are shown. The data and the simulation show a similar trend.
Note that the errors given are correlated, since the data in one bin are a subset of
the data in the next. The solid line corresponds to the $(1-p)$ dependence discussed
above. The linear relationship between $p$ and $\Delta\varphi_d$ is estimated from the
phase-space distribution of particles in RQMD, including effects from reconstruction
efficiency and background tracks. For larger angles, both data and the RQMD results
lie consistently below the line, which indicates that effects from resonance decays
are important.

Due to the influence of the magnetic field the positive and negative particles will
have different azimuthal acceptance. The $\Delta\varphi_d$ study in Fig. 2 thus
selects partly non-overlapping regions of phase space for positive and negative
particles. A remedy for this is to use the reconstructed $\varphi$-angle for each
particle $\varphi_r$, i.e. the azimuthal direction of the particle at the primary
vertex, before it is deflected by the magnetic field. Figure 3a shows the acceptance
in transverse momentum and $\varphi_r$ for positive and negative particles. By
choosing the azimuthal interval $\Delta\varphi_r$ symmetrically around the center of
the acceptance, a better phase space overlap is achieved for small azimuthal
intervals. In Fig.~3b, $v(Q)$, for the 10 \% most central events, is displayed as a
function of $\Delta\varphi_r$. The $(1-p)$ dependence, which is no longer linear, is
given by the solid curve. Again data and the RQMD results show a similar trend, but
the deviations from the curve are larger in this representation, indicating that an
overlap in phase space is of importance. For large values of $\Delta\varphi$, the
acceptance approaches the limits determined by the boundaries of the tracking arm, and
the two representations are essentially the same.

The effects of the detector inefficiency and background tracks not assigned the
correct charge have been investigated in a Monte Carlo simulation. It is assumed that
the inefficiency independently removes positive and negative particles with the same
probability, and that the background consists of uncorrelated positive and negative
particles. The reconstruction efficiency and the amount of background have been
determined from the RQMD and GEANT simulations discussed earlier. Both the
inefficiency and the background contribution have the effect of diluting the signal
and pushing the value of $v(Q)$ closer to 1. The dilution due to these effects can be
treated as an experimental systematic error, estimated from the simulations, setting a
lower limit on $v(Q)$. For the net charge fluctuations in the region -0.35 $< \eta <$
0.35, $p_T >$ 0.2 GeV/c, $\Delta\varphi = \pi/2$,
\begin{equation}
v(Q) = 0.965 \pm 0.007 (stat.) - 0.019 (syst.) \label{e6}
\end{equation}
is obtained for the 10 \% most central events. A linear extrapolation of this value to
full azimuthal coverage gives a value of v(Q) in the range 0.78 - 0.86, in qualitative
agreement with calculations from a hadronic gas. For comparison, it would be desirable
to have a Monte Carlo model for the QGP which exhibits the predicted reduction in the
charge fluctuations and which could be used to study the sensitivity of the method for
limited acceptance.

To summarize, we have shown that the data behave in an almost stochastic manner. There
are also clear indications that effects from hadronic decays are seen; the data are in
good agreement with RQMD calculations, which includes the effects of global charge
conservations as well as neutral hadronic resonance decays. Furthermore, the data show
no centrality dependence, which is in contradiction to the expectations from a
Quark-Gluon Plasma scenario. We have clearly demonstrated that the fluctuations of the
charge ratio $v(R)$ and of the net charge $v(Q)$ are well understood in a stochastic
model. We, however, advise against the usage of the proposed R variable~\cite{Jeon},
since it unnecessarily complicates the evaluation of the fluctuations, and the
intrinsic decrease of $v(R)$, as a function of centrality, can be mistaken as a
'plasma fingerprint'. 
The measured value of $v(Q) =$ 0.965 is far from the value
predicted for a plasma. Even extrapolating the linear trend seen in the data in Fig. 2
to full azimuthal coverage, renders values of the fluctuations, which are far above
the proposed values. With the caveat of our limited acceptance in rapidity, these
results clearly indicate either the absence of a plasma or that the proposed signal
does not survive the transition back to hadronic matter.

%
%
%
%
%
%
%


We thank the staff of the Collider-Accelerator and Physics Departments at
BNL for their vital contributions.  We acknowledge support from the
Department of Energy and NSF (U.S.A.), Monbu-sho and STA (Japan), RAS,
RMAE, and RMS (Russia), BMBF, DAAD, and AvH (Germany), VR and KAW
(Sweden), MIST and NSERC (Canada), CNPq and FAPESP (Brazil), IN2P3/CNRS
(France), DAE and DST (India), KRF and CHEP (Korea), the U.S. CRDF for 
the FSU, and the US-Israel BSF.







\begin{figure}
\vspace{3cm}
\centerline{\epsfig{file=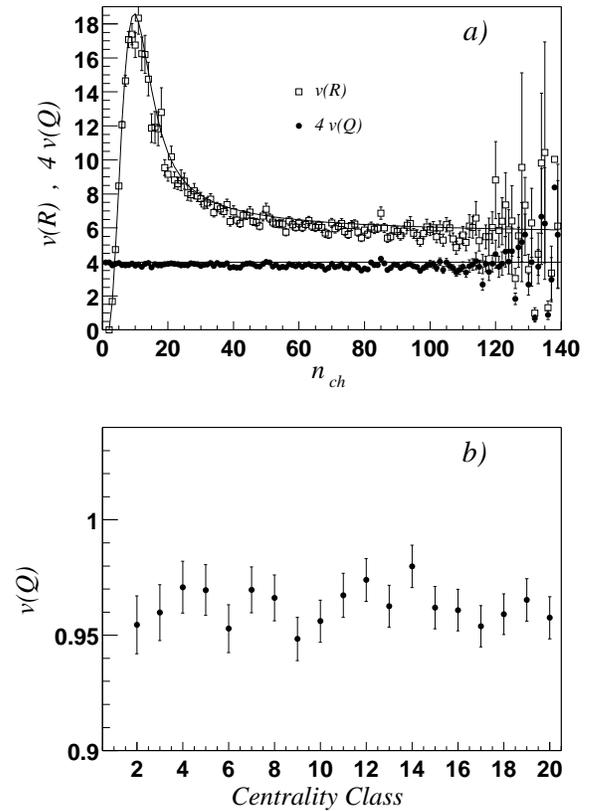,width=1.0\linewidth}}
\caption[]{
a) The normalized variances $v(Q)$ and $v(R)$ as functions of $n_{ch}$,
together with curves for stochastic behaviour.  b) The normalized variance
$v(Q)$ for different centrality classes, as described in the text.
}
\label{fig:1}
\end{figure}
   
\newpage
\begin{figure}
\centerline{\epsfig{file=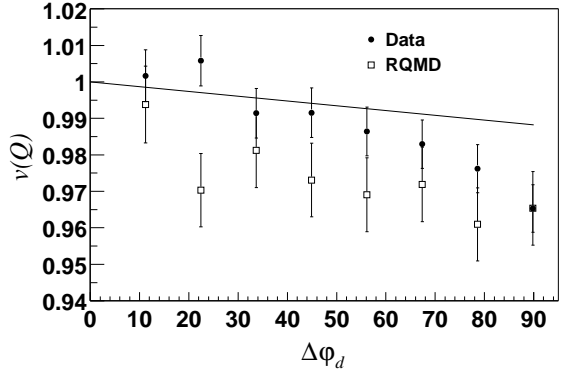,width=1.0\linewidth}}
\caption[]{
$v(Q)$, for central events, as a function of the azimuthal coverage of the
detector for data and events simulated with the RQMD model. The solid line
shows the expected reduction in $v(Q)$ in the stochastic scenario when
global charge conservation is taken into account. (Angles in degrees.)}
\label{fig:2}
\end{figure}


\begin{figure}
\vspace{0.1cm}
\centerline{\epsfig{file=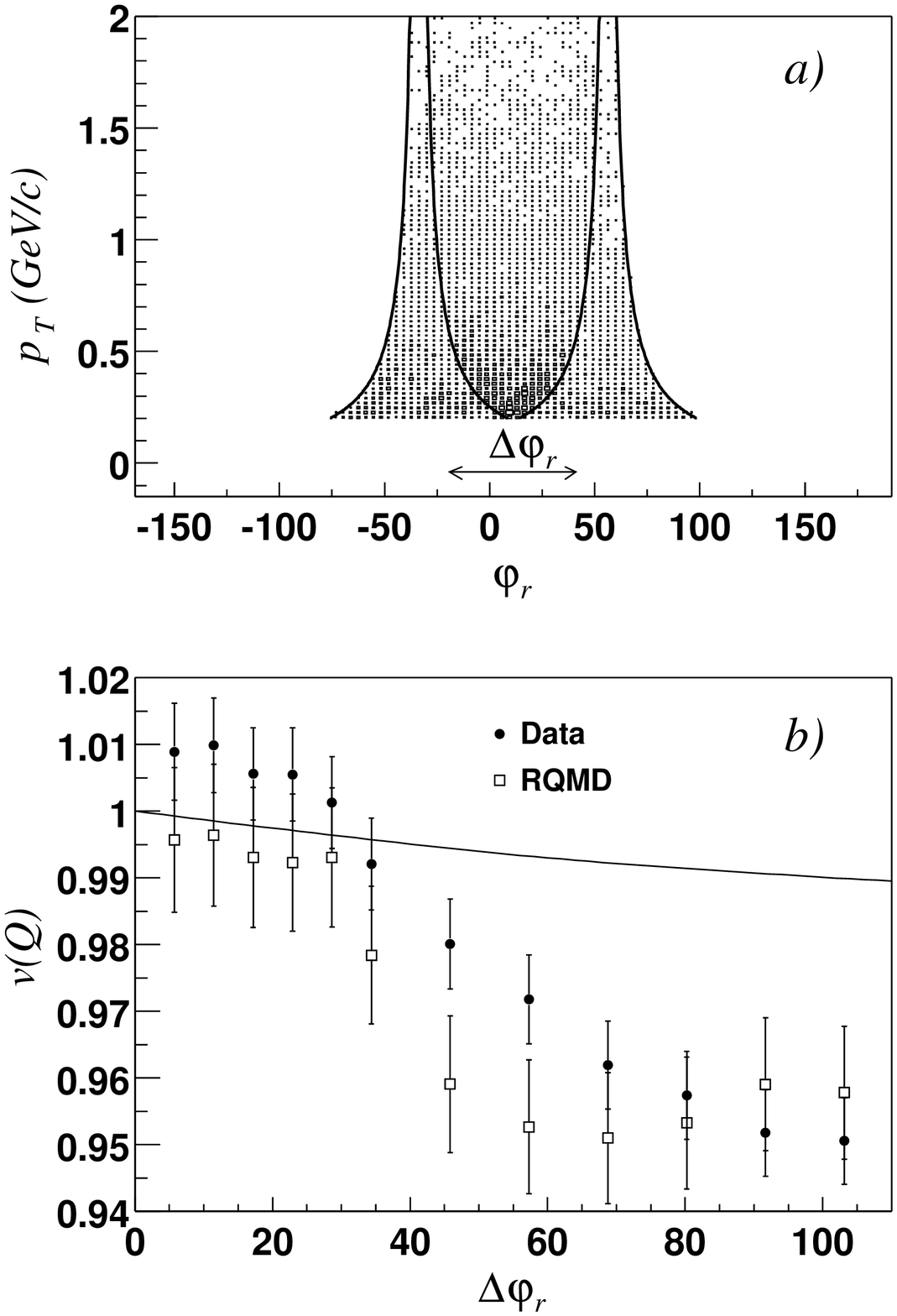,width=1.0\linewidth}}
\caption[]{
a) The acceptance in reconstructed transverse momentum $p_T$ and azimuthal
angle $\varphi_{r}$. The solid curves indicate the acceptance bands for
positive and negative particles, respectively.
b) The effect on $v(Q)$ of varying the acceptance in $\varphi_{r}$. The
solid curve shows the expectation from global charge conservation. The 10
\% most central events are used for data and RQMD. (Angles in degrees.)
}
\label{fig:3}
\end{figure}




\end{multicols}    


\begin{references}

\bibitem[*]{Deceased}Deceased     

\bibitem[\dagger]{non-par} Not a participating Institution.

\bibitem{PHENIX} PHENIX Collaboration, D. P. Morrison, {\it{et al.}},
Nucl. Phys. {\bf{A638}}, 565c (1998).

\bibitem{QGP} see e.g. J.-P. Blaizot,  Nucl. Phys. {\bf{A661}}, 3c (1999).

\bibitem{Jeon} S. Jeon and V. Koch, Phys. Rev. Lett. {\bf{85}}, 2076 (2000);
  M. Bleicher, S. Jeon and V. Koch, Phys. Rev. {\bf{C62}}, 061902(R) (2000).

\bibitem{Asakawa} M. Asakawa, U. Heinz and B M\"uller, Phys. Rev. Lett.
{\bf{85}}, 2072 (2000).

\bibitem{Heiselberg} H. Heiselberg and A. D. Jackson, Phys. Rev.
{\bf{C63}}, 064904 (2001).

\bibitem {Shuryak} E. V. Shuryak and M. A. Stephanov, Phys.Rev.
{\bf{C63}}, 064903 (2001).

\bibitem {Fialkowski} K. Fialkowski and R. Wit, Europhys. Lett. {\bf{55}}
(2), 184 (2001).

\bibitem{Jeon2} S. Jeon and V. Koch, Phys. Rev. Lett. {\bf{83}}, 5435 (1999).

\bibitem{Dumitru} A. Dumitru and R. D. Pisarski, Phys. Lett. {\bf{B504}},
282 (2001).

\bibitem{Wolf} see e.g. E.A. De Wolf, I.M. Dremin and W. Kittel, Phys.
Rep. {\bf{270}}, 48 (1996).

\bibitem {Mult} PHENIX Collaboration, K. Adcox, {\it{et al.}}, Phys. Rev.
Lett. {\bf{86}}, 3500 (2001).

\bibitem {general} V. G. Riabov, Nucl. Instr. and Methods {\bf{A 419}},
363 (1998).

\bibitem {jeff} J. T. Mitchell et al., nucl-ex/0201013, accepted for
publication in Nuclear Instr. and Methods A.

\bibitem {GEANT} GEANT 3.2.1, CERN program library.

\bibitem {RQMD} H. Sorge, Phys. Rev. {\bf{C 52}} 3291 (1995).


\end{references}
\end{document}